\documentclass[3p,preprint,twocolumn,authoryear]{elsarticle_mod} 

\usepackage{t1enc}
\usepackage{verbatim}
\usepackage{amsfonts,amssymb}
\usepackage{url}
\usepackage{datetime}
\usepackage{ragged2e}

\begin{document}
\def\elsA/{{\sffamily\itshape\mdseries elsA}}
\def\elsAPython/{{\sffamily\upshape\mdseries Python--\elsA/}}
\def\elsAPythonAPI/{{\sffamily\upshape\mdseries Python--\elsA/\_API}}
\def\PyGelsA/{{\sffamily\upshape\mdseries PyG\textit{elsA}}}
\def\DoE/{{\textsc{d}o\textsc{e}}}
\def\AoA/{{\textsc{a}o\textsc{a}}}
\newcommand*{\sigle}[1]{{\protect\scshape\MakeLowercase{#1}}}
\newcommand*{\comput}[1]{{\upshape\ttfamily\small #1}}
\newcommand*{\computn}[1]{{\upshape\ttfamily #1}}
\newcommand*{\cmdopt}[1]{{-}{-}\comput{#1}}
\renewcommand{\figurename}{Fig.}

\makeatletter
\newcommand{\verbatimfont}[1]{\renewcommand{\verbatim@font}{\ttfamily#1}}
\makeatother

\def\UrlBreaks{\do\/\do-}

\verbatimfont{\small}
\setcounter{tocdepth}{2}

\begin{frontmatter}

\title{The Python user interface of the \elsA/ \textsc{cfd} software:\\a coupling framework for external steering layers}
\author[on]{Marc Lazareff\corref{cor1}}
\address[on]{ONERA BP72 - 29 avenue de la Division Leclerc FR-92322 CHATILLON CEDEX}
\ead{marc.lazareff@onera.fr}
\date{\today}

\begin{abstract}

The \elsAPython/ user interface of the \elsA/ \sigle{CFD}
(Computational Fluid Dynamics) software has been developed to allow
users to specify simulations with confidence, through a global context
of \emph{description objects} grouped inside \emph{scripts}. The
software main features are generated documentation, context checking
and completion, and helpful error management. Further developments
have used this foundation as a coupling framework, allowing (thanks to
the descriptive approach) the coupling of external algorithms with the
\sigle{CFD} solver in a simple and abstract way, leading to more
success in complex simulations. Along with the description of the
technical part of the interface, we try to gather the salient points
pertaining to the psychological viewpoint of user experience
(\sigle{UX}). We point out the differences between user interfaces and
pure data management systems such as \sigle{CGNS}.

\end{abstract}

\end{frontmatter}

\section{Introduction}

\elsA/ (\url{http://elsa.onera.fr/}) is a large
\sigle{CFD} software for research and industry, mainly used in
aerospace design, for both internal and external flow. It has been
described before \citep{Gazaix2002,Cambier2011}.

Here we will be interested in the extension of the \elsAPython/ user
interface to additional software, for performing simulations not only
at a few discrete specified workflow conditions (e.g. given Mach,
Reynolds \ldots) but on a domain of variation, i.e. a Design of
Experiment space (\DoE/), with several algorithmic layers added above
the \sigle{CFD} solver.

The first step is to automate the spanning of a (given) large number
of specified points covering the \DoE/. The second one is to add a
stabilization layer for prolongation through unstable \DoE/ zones
(e.g. when the flow conditions lead to separation). The third one is
to use the resulting ``stabilized spanning'' algorithm as a provider
of \emph{observable quantities} for a sparse polynomial interpolator,
driving the simulation and selecting the \DoE/ points to compute. If
successful, this composite algorithm provides (for the chosen
observable) an efficient global representation (response surface)
allowing further studies on the \DoE/, at low cost; for example
stochastic analysis using the Monte-Carlo method on the response
surface would be a fourth step.

The specific point of interest of our approach is that, for this
complex composite algorithm, we manage to keep the user interface
simple and uniform, using the same concepts as for the base interface.
Sections \ref{orient}-\ref{statdyn} explain these concepts; sections
\ref{doc}\&\ref{pygelsa} deal with documentation, error messages, and
the \sigle{GUI} (Graphical User Interface); finally, sections
\ref{datanet}-\ref{doe} deal with high-level operations and how we use
them for \DoE/ spanning and analysis.

\enlargethispage{1cm}

This paper includes a number of elements from the \sigle{CFD}
application domain, and software considerations, along with user
interface concepts; these concepts (and most of the software
implementation) would be applicable to many different domains. Not
included here are considerations for field-like data, for which the
\sigle{CFD} General Notation System standard (\sigle{CGNS})
is better suited\footnote{But \sigle{CGNS} is not a user
interface system, lacking even basic user-oriented checks}.

\section{Interface development orientation}
\label{orient}

\subsection{User interaction model}

In the development of the \elsAPython/ user interface, we strived to
reach an equilibrium between power and usability, both on the user
side (running simulations) and on the developer side (adding
functionality); also, we have tried to avoid a steep user learning
curve.

The \elsAPython/ user interface has been used since June 2000 with no
change in its base principles (see section \ref{base}), and thus maybe
it is not too far off its target demography. The basic interface is
readily extended with additional classes allowing external algorithmic
layers, see \ref{products}, to complement the \sigle{CFD} solver coded
in the \elsA/ kernel.

The \elsAPython/ interface is built on a \emph{declarative} user
interaction model \citep{declarative}, whereas similar software
\citep{Gerhold2008} uses an \emph{imperative} model
\citep{imperative}. The scripts describing the \elsA/ \sigle{CFD}
simulations use context sharing to avoid data duplication; and all the
tedious details which are possibly automated are kept hidden in the
behaviour of the description classes.

The Python language has been a great help in building successive
abstract layers. We generally try to go with the Principle of Least
Astonishment \citep{Seebach2001,Ronacher2011}, and to facilitate an
object-oriented and \emph{functional}
\citep{functional,Backus1978b,Hudak1989} programming style, here in
user scripts.

The (functional) Reverse Polish Lisp (\sigle{RPL}) language of some
Hewlett-Packard handheld calculators, with its three language levels
(User \sigle{RPL}, System \sigle{RPL}, compiled code) \citep{RPLwiki},
has been the inspiration for an intermediate language where, as for
System \sigle{RPL}, a faster checks-less interpreted code -- usually
generated on the fly from the User level -- may be directly accessed
by experts.

\medskip

In \elsAPython/, user interaction is mainly through argument-less
method calls; this is possible because the would-be arguments are
already known as attribute values of the method's owner object; these
values may originate:
\begin{enumerate}[a)]

\item from user specification, using the owner object's \comput{set()}
  method, see \ref{methods_desc};

\item from context rules -- \comput{set()} terminator in
  context-dependent default rules for the owner object -- possibly
  involving dependencies on objects other than the method's owner,
  see \ref{context_manage};

\item using the \comput{set()} method in the behaviour logic of other
  objects, possibly also involving context rules (e.g. in coupled
  problems), see \ref{sfddmd}.

\end{enumerate}

The user interaction model thus includes an explicit part (method
calls in the user script) and an implicit part (context rules and
other class-specific object behaviour), providing
automatically-defined values. The latter part is essential, allowing
for a lighter work burden and thus better efficiency and security in
problem solving. The potential danger, regarding user confidence, of
automatically-defined attribute values is alleviated by the
\emph{display} (\comput{view} method, see \ref{methods_cntx}) and
\emph{introspection} features of the interface (\comput{show\_origin}
method, see \ref{contdefs}).

\subsection{Base principles, user \& developer sides}
\label{base}

The base elements and principles used in the development of this
interface are:
\smallskip

\noindent\emph{For the user (U) side:}

  \begin{enumerate}[U1.]

  \item declarative model: the problem definition is made up of
    definitions of the simulation parameters; building the adequate
    solver (acting) is left to a \emph{factory}, lower down in the C++
    \elsA/ kernel; this separation of roles shields the interface from
    modifications in the kernel implementation details;

  \item object-oriented model: the simulation parameters (elements of
    the mathematical model) are here represented as attributes of
    \emph{description} classes; the whole simulation \emph{script} (a
    \emph{description}s container) is also a class instance;
    \emph{script}s include no data, only \emph{description} creations
    and a few operations, like \comput{<desc>.set(<attribute>,
      <value>)}, \comput{<desc>.check()} or \comput{<desc>.compute()};

  \item implicit hierarchy: the whole simulation data structure is a
    shallow (four-level) tree of scripts (possibly nested),
    descriptions, attributes, and values (see \ref{static_tree}); this
    tree is not explicitly referenced in interface use\footnote{Unlike
      a \sigle{CGNS} tree -- managed in \elsA/ through \comput{py\sigle{CGNS}}
      -- which large depth is especially apparent when it is accessed
      through the \comput{\sigle{CGNS}view} graphical interface.};

  \item static checks (attribute name and value type/definition
    domain) are performed by default on attributes , see
    \ref{statdefs};

  \item dynamic (contextual) checks are performed on demand to ensure
    the coherency of the problem description; a
    \emph{context}\footnote{Contextual (influence/dependence)
      relations introduce a different graph than the above shallow
      tree; this second graph, traversed by the \comput{check()}
      method, see \ref{methods_cntx}, may be deep (have a large number
      of levels), see \ref{dynamic}, and is not always a tree, see
      \ref{context_manage}.} is an instance of either a
    \emph{description} class or a \emph{script} class, both owning a
    \comput{check()} method.

  \item context-dependent (contextual) default values allow tailoring
    the software to different trades, see \ref{contdefs}; the origin
    rule for default values may be traced back through the
    \comput{show\_origin()} method;

  \item keep the user in charge: no automatic modification of his/her
    input; conflicts may only be resolved by the user, upon
    notification by the interface;

  \item the interface may be used as a standalone checking tool, with
    no \sigle{CFD} kernel needed;

  \item the user interface scope is limited to scalar data: mesh and
    field data (e.g. files) is referenced without contents checking.

  \end{enumerate}

\pagebreak[2]

\noindent\emph{For the developer (D) side:}

  \begin{enumerate}[D1.]

  \item the \sigle{API} (Application Programming Interface) between
    the user interface and the \elsA/ kernel is based on three basic
    types -- \comput{float} (floating-point), \comput{int}
    (integer) and \comput{string} -- and a few \emph{methods} and
    \emph{functions}, see \ref{methfunc};

  \item the evolving part of the interface is mainly described through
    resource files, see section \ref{statdyn}, from which the
    documentation skeleton ({\LaTeX} commands, see \ref{urm}) is
    automatically generated in a coherent way;

  \item functionality may be added through ``products'', see
    \ref{products}, the Python code for which is simply dropped in the
    adequate place, defining additional \emph{description} classes;

  \item obsolete elements of the interface, see \ref{metadata}, may be
    re-activated for version comparison.
 
  \end{enumerate}

\subsection{Naming things}
\label{naming}

One important aspect of the interface -- for which there are no exact
rules and only a few general principles -- is the \emph{naming} of the
various interface elements (especially attributes and their values),
see \ref{naming}. This aspect is more in the realm of linguistics, and
must take into account the culture of the users.

Naming is recognized as a hard problem in software
\citep{Deissenboeck2006}, and is all the more important when dealing
with user interaction. In programming, users better understand long,
self-explanatory, names \citep{Binkley2009a}, but ``The nonsense words
did surprisingly well \ldots distinctive names are helpful even when
they are not meaningful'' \citep{Shneiderman1997}.

It seems then that there is no symmetry between understanding which
concept is behind a name and, inversely, remembering which name
corresponds to the sought concept. It is probably better not to use
long names which may differ only by one or two characters -- better
use shorter names where the differences stands out -- or to use names
with easily confused characters \citep{Kupferschmid2009}. Also, we
have chosen the underscore style, e.g. \comput{global\_timestep}, for
the user interface rather than ``camel casing'',
e.g. \comput{GlobalTimestep}, although the latter seems more popular
with programmers \citep{Binkley2009b}.

A balance must somehow be found, for users to be able both to read and
write scripts without reaching for the documentation at every
step. This part of the interface management is probably the most
involved with language considerations, and the most dependent on user
cultural background (e.g. when reusing variable names from equations).

\section{\elsAPython/ description language}
\label{elsAPython}

\elsAPython/ is built upon the Python language to create a standard of
creation and modification of description (attributes container) and
script (descriptions container) objects. Only the main elements of the
language will be described here, hoping they give a taste of the
chosen \emph{user interaction model}. Both for the text-mode and
graphical-mode interfaces, we have tried to respect the design
principles of \citep{Shneiderman1997}, with \citep{Clarke1986} also as
a (more abstract) background.

From then on, the \comput{<>} brackets will be used with general
meaning of ``realization'' (instance, value, \ldots) of the enclosed
symbol (class, attribute, \ldots).

\subsection{Description object creation}

Creating an instance of the \comput{<desc>} description class is performed by:

\comput{<name> = <desc>(name='<name>')}

where \comput{<name>} (without quotes) is the canonical Python
reference to the created object and \comput{'<name>'} a text
identifier using the same characters. This identifier is used both for
allowing forward references (to not yet created objects) and for
out-of-memory references, either between computational nodes without
shared memory or in databases (for accessing objects directly, not
through property search), see \ref{database}.

Notice: simple test scripts may forgo the \comput{<name>=} argument,
thanks to an automatic canonical name search feature, but this is very
costly for large scripts and is precluded in industrial cases.

\subsection{Script object creation}

The \comput{script} class is derived from the Python \comput{module}
class, to which context-oriented behaviour is added.

A \comput{<script>} (\comput{script} instance) is automatically
created in-memory when loading a script file (containing Python code)
from the \elsA/ command-line; this is the most common way a
\comput{<script>} object is created.

Internally, this creation is performed through the \elsAPython/
\comput{load()} function, of which the simplest call form is:

\comput{<script> = load(<script\_file\_name>)}

More scripts may be loaded from the main (\comput{root}) script,
\emph{script} objects may thus be nested.

\subsection{Global tree structure}
\label{static_tree}

Building a problem description in \elsAPython/ implicitly creates a
tree, with the enclosing (top-level) \emph{script} object as
\comput{root}:

\begin{itemize}
\setlength\itemsep{0ex}

\item \emph{script}s reference (include) \emph{description}
  constructors (and other \emph{method}, or \emph{function}, calls),
  but own no \emph{attributes};

\item a \emph{script} may reference (include) other \emph{script}s,
  through \comput{load()} calls;

\item a \emph{description} may reference other \emph{description}s,
  through \comput{attach()} calls, e.g. \comput{cfd1.attach(mod1,
    num1)};

\item \emph{description}s own (contain) \emph{attribute}s;

\item \emph{attribute}s have \emph{value}s;

\item \emph{value}s may be \comput{<float>}, \comput{<int>},
  \comput{<string>}, or \comput{None} (meaning ``no value'', for all
  types).

\end{itemize}

Calls to the \comput{check()} method of \emph{script}s and
\emph{description}s (both being \emph{contexts}, see
\ref{methods_cntx}) will transitively traverse the referenced objects
(included \emph{script}s, included or attached \emph{description}s).

\subsection{Methods and functions}
\label{methfunc}

We describe below a few methods of description classes, see
\ref{methods_desc}, and of contexts, see \ref{methods_cntx};
additionally, the \comput{close()} function call will end the
processing of a script.

\subsubsection{Specific methods of description classes}
\label{methods_desc}

\begin{itemize}
\setlength\itemsep{0ex}

\item \comput{<desc>.set(<attr>, <valu>)}: define \comput{valu} as the
  value of the \comput{attr} attribute of the \comput{<desc>}
  description class instance;

\item \comput{<desc>.get(<attr>)}: return the value of the
  \comput{attr} attribute of the \comput{<desc>} description class
  instance;

\item \comput{<desc>.compute()}: start the algorithm managed by
  the \comput{desc} description class.

\end{itemize}

\subsubsection{Methods of contexts (descriptions and scripts)}
\label{methods_cntx}

Description classes and script classes are both contexts: they all
inherit from the \comput{context} class, and thus share a number of
methods, which are shown here (in the simplest syntax).

\begin{itemize}
\setlength\itemsep{0ex}

\item \comput{<context>.check()}: perform contextual checking (using
  influence, dependency and contextual default rules) on the
  \comput{<context>} instance; this method is the defining feature of
  a context;

\item \comput{<context>.view()}: display a compact (using
  macro-attributes) representation of the \comput{<context>} instance,
  masking all non-coherent attribute values;

\item \comput{<context>.copy(name=<name>)}: return a copy of the
  \comput{<context>} instance, with the \comput{<name>} identifier;

\item \comput{<context>.dump()}: dump a compact representation of the
  \comput{<context>} instance to a file; dumped scripts may not
  include explicit control structures (loops and tests), which
  should be managed by description classes.

\end{itemize}

\subsection{Abstraction \& flexibility: redirections}
\label{redirect}

For better abstraction and flexibility, some functions and methods
are redirections to lower levels:

\begin{itemize}
\setlength\itemsep{0ex}

\item function to \comput{root} script method;

\item script method to \comput{rootboot} description method.

\end{itemize}

as explained below in \ref{functions}, \ref{methods_scpt}. These
redirections are most powerful when coupling with external layers, see
\ref{products}.

\subsubsection{Functions redirected to \comput{root} script methods}
\label{functions}

The \comput{compute}, \comput{extract}, \comput{check} and
\comput{dump} functions are indirections to the same-named methods of
the top-level \comput{root} script object. Moreover, the
\cmdopt{check} and \cmdopt{dump} command-line options
(\sigle{CLO}) trigger the corresponding function call on an explicit
\comput{close()} call or on natural termination of the \comput{root}
script, while the \cmdopt{strict} option triggers a
\comput{check()} call before the \comput{compute()} call.

\subsubsection{Script methods redirected to the \comput{boot} description}
\label{methods_scpt}

The \comput{compute} (start solver) and \comput{extract} (build the
specified output representation) methods of the current script object
are indirections to the same-named methods of the current
\comput{boot} description object, see \ref{rootboot}.

\section{Static and dynamic behaviour}
\label{statdyn}

New description classes do appear from time to time, but the main
mechanism for the evolution of the \elsAPython/ interface is the
addition of new attributes to existing classes. This is performed by
augmenting two resource files, one for static definitions and the
other for dynamic (context-related) ones.

All resource files (data defining the class attributes and various
contextual rules) are structured using varying combinations of the
\comput{dictionary} and \comput{list} Python types; these
\emph{varying} combinations are allowed (without additional coding)
because Python is a dynamically typed language.

Rather than using a uniform formal schema, different \emph{ad hoc}
grammars are used here for each kind of definition, abusing the expert
(if dated) advice: ``As far as we were aware, we simply made up the
language as we went along.'' - John Backus, Developer of Fortran
(1957) and inventor of \sigle{BNF} (1959) \citep{Backus1978a}.

We thus conveniently ``forget'' the invention of the more formal
Backus Normal (later Naur) Form (\sigle{BNF}), and use a homebrewed
grammar; this non-formal character introduces a degree of
incompleteness, which has not introduced problems so far for the
\sigle{CFD} target application field; a few special cases of rule
definitions are treated using \comput{lambda} expressions (anonymous
functions for in-lining rule code), with also some regular expression
matching in lieu of plain comparison on \comput{<string>} values.

Using only the basic \comput{dictionary} and \comput{list} Python
types for these homebrewed grammar definitions makes the corresponding
resource files compact and readable, so that developers may actually
augment them; they are also very fast to parse. Again referencing
(albeit indirectly) John Backus: ``Because the customers of the 704
were primarily scientists and mathematicians, the language would focus
on allowing programmers to write their formulas in a reasonably
natural notation.''  \citep{Aiken2007}.

\subsection{Definitions for static behaviour}
\label{statdefs}

The definitions for static behaviour (i.e. excluding contextual
checks) are centralized in a global file, as a Python module. They
include for each attribute:

\begin{itemize}
\setlength\itemsep{0ex}

\item a short descriptive text;

\item a type (float, integer or string) for the attribute value;

\item one or several checking methods, further restricting the
  definition domain (e.g. \comput{<float>} type but further restricted
  to $\mathbb{R}^{*+}$);

\item a list of default value mechanisms, among: static (value),
  dynamic (reference to rule), \comput{None};

\item optionally, restrictions to modifications of the attribute
  value, e.g. interface only (not user).

\end{itemize}

Example:
\newline
\comput{'phymod':["""fluid model""",['S','I'], \{'euler':0,'nslam':1,'nstur':2\}, [CNTX\_DEFV,None]]}

These definitions are rendered in plain English (along with the
context-related items, see \ref{dynamic}) by the \comput{man()} function
of the interface, see \ref{intman}.
Moreover, a number of entries may be defined for each class, defining
additional metadata, see \ref{metadata}.

Part of the attributes definitions (for example inheritance,
see \ref{metadata}) is only finalized at runtime. It is then used to
define class singletons, managing the attribute definitions for all
the instances of a given class. This allows caching (across same-class
instances) the internal representation of the definitions.

\subsubsection{Attribute definition details}

\paragraph{Description chain}

The description chain for each attribute is used by the \comput{man()}
function, the popups in the \PyGelsA/ \sigle{GUI}, see section
\ref{pygelsa}, and for generating the skeleton of the User's Reference
Manual, see section \ref{doc}.

\paragraph{Type and definition domain method(s)}

The basic type (float, integer or string) of the values of each
attribute is complemented with a list of allowed of values, a
(registered) checking method for the definition domain
(e.g. ``strictly positive'' or more complex), or a combination of
both. When the type changes between the interface and the kernel, the
list of allowed values is replaced with a conversion dictionary.

\paragraph{Default values}

The ``default values'' item may be an explicit value, a reference to
the (possibly converted) kernel default value, a reference to a
(possibly non-existent) context-dependent (contextual) default value,
or a list of such items (e.g. contextual, then ``static'' safe
value). The single \comput{None} default value means that the
attribute has no default value; if moreover the attribute is defined
as always required, context checks will return \comput{False} unless
it has been defined by the user.

\subsubsection{Macro-attributes}

To give structure to the large number of attributes of some classes,
they may be grouped into ``macro-attributes'', meaning named lists of
attributes. Macro-attributes are list-valued and may be managed
through \comput{set()}, \comput{get()}, \comput{view()} and generally
all methods of the description classes meant for plain attributes
(their ``atoms'').

A macro-attribute may have several versions with differing lengths,
e.g. \comput{conservative} is always the name of the list of the
attributes for the conservative variables, whatever the current number
of equations. Specific versions (as declared in the resource file) are
internally named e.g. \comput{conservative*05},
\comput{conservative*06} \ldots but on the user side, for input and
output, plain \comput{conservative} is used. This feature is not
guaranteed to always provide as easy a grouping of related attributes,
but it is quite useful.

In the \sigle{GUI}, see section \ref{pygelsa}, the widgets for
macro-attributes appear as foldable groups of ``atom''
widgets. Folding is automatic on failed influence/dependency rules
(meaningless macro-attribute).

\subsubsection{Attributes metadata}
\label{metadata}

Additional attributes metadata are structured in class-wide lists for:

\begin{itemize}
\setlength\itemsep{0ex}

\item always-required attributes (whose value must be defined);

\item obsolete attribute and values, and possibly their current
  replacement\newline (\cmdopt{allow\_obsolete} \sigle{CLO});

\item attribute and values which may be filtered out (\cmdopt{filter}
  \sigle{CLO});

\item not yet documented attributes and values (\cmdopt{unlock} \sigle{CLO});

\item inheritance (with possible modifications) of attribute
  definitions from other classes.

\end{itemize}

\subsection{Definitions for dynamic behaviour}
\label{dynamic}

The definitions for dynamic behaviour are the most important part of
the base interface, allowing for context management and thus more
abstract problem specification. Context management allows the software
to behave differently, according either to a few trade-specific
indications, see \ref{contdefs}, or to the current global solving
operator (in case of coupling), see \ref{sfddmd}.

\subsubsection{Context management}
\label{context_manage}

The definitions for dynamic behaviour (i.e. rules for context
management) are centralized in a global resource file, as a Python
module; the inference engine (for applying these rules) is defined in
a separate module. Dynamic behaviour is represented here by the
context-dependent features of the interface: influence and dependency
rules, and contextual default rules, completed by the traversal of
relations created through \comput{attach()} calls and possibly by the
hierarchical structure of scripts (through inclusion).

Each problem description thus introduces a realization of a dynamic
structure, whose nodes are linked by contextual influence and
dependence relations; this structure is a Directed Acyclic Graph
(\sigle{DAG}), and not a true tree, because a node may have more than
one parent (several influence rules may lead to the same attribute
being required). The ``Acyclic'' word is important (also valid for a
tree, of course), because we need to ensure that no rules in the
grammar lead to circular checking. The ``dynamic'' qualifier comes
both from the influence and dependence rules and from the
context-dependent default values, leading to (possibly) different
graph realizations for different user inputs, contrary to the static
layout of a \sigle{CGNS} tree.

Default values are sought when a rules terminator (an end rule)
specifies that an attribute value must be defined, and no such value
exists; \emph{contextual} default rules, see \ref{contdefs}, depend on
the current state of the descriptions context. The
\comput{<desc>.get\_or\_deft(<attr>)} method call returns the
best-effort result for the value of the \comput{<desc>.<attr>}
attribute -- using available context state (attribute values) and
rules -- or \comput{None} if no default value may be computed.

\begin{figure}[!h]
\centerline{\includegraphics[width=1.05\columnwidth]{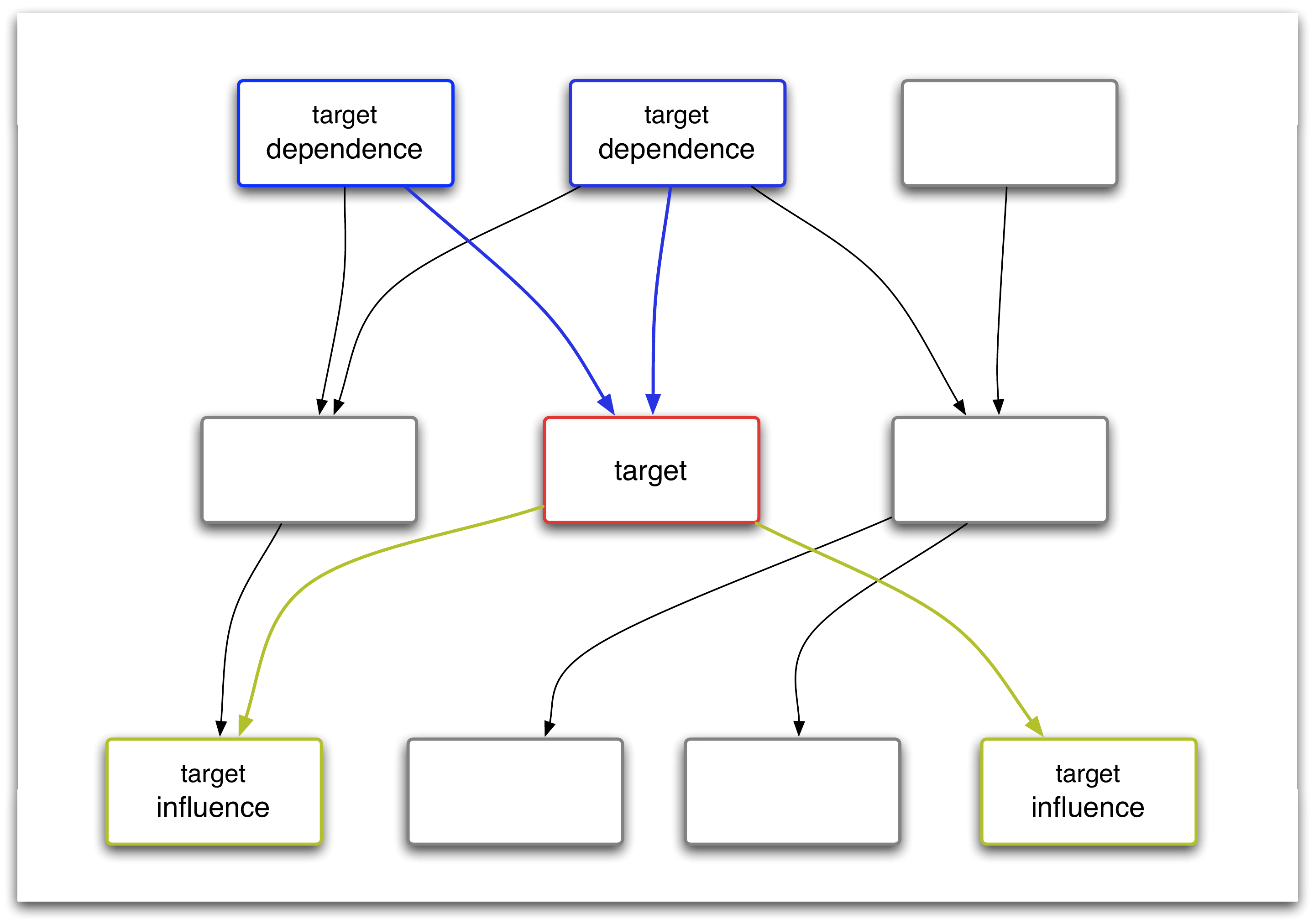}}
\caption{\label{tree_dep_inf}Contextual \sigle{DAG} information flow,
  through influence and dependency.}
\end{figure}

\begin{figure}[!h]
\centerline{\includegraphics[width=1.05\columnwidth]{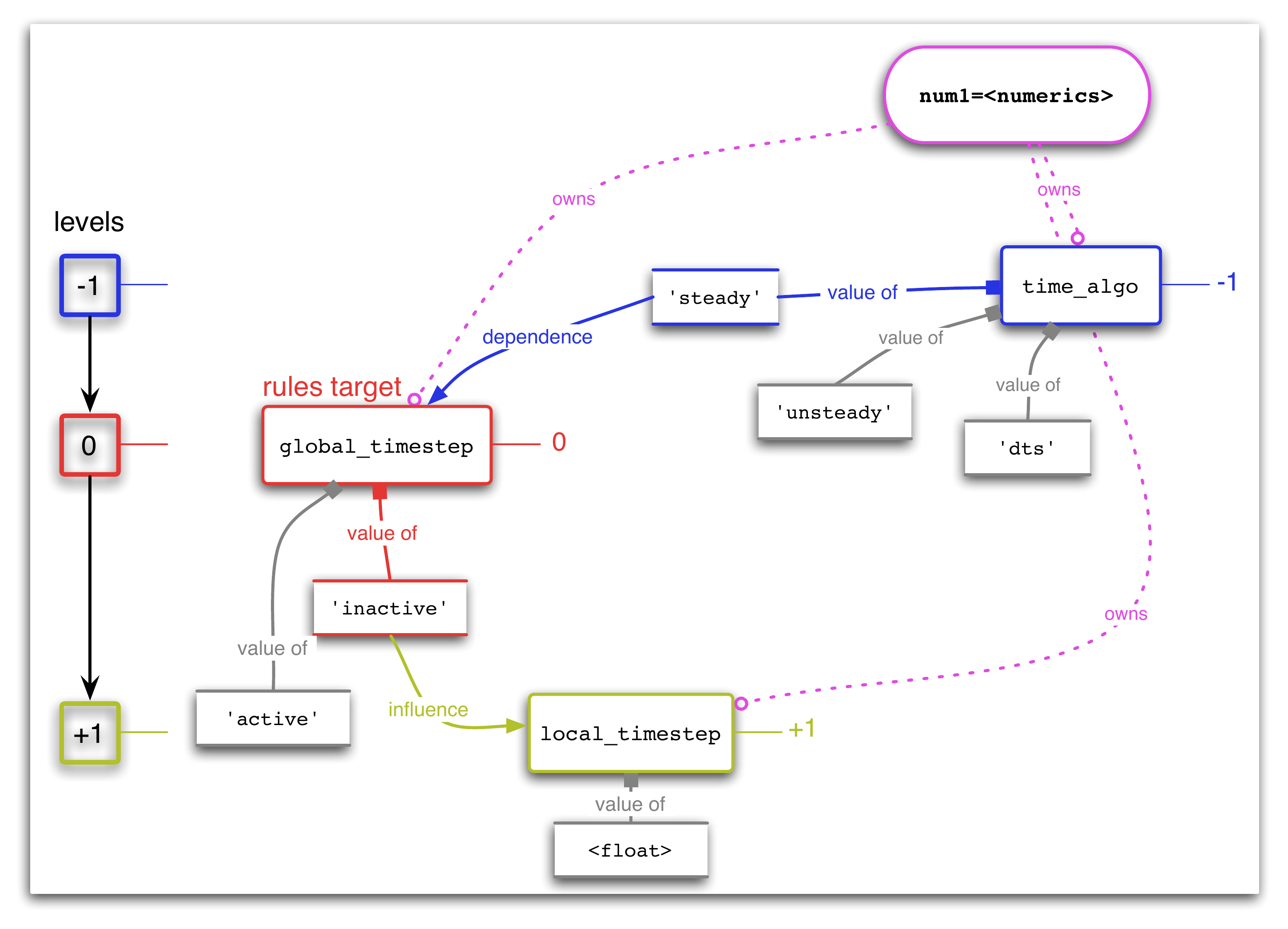}}
\caption{\label{local_tree}Local rules evaluation (one attribute).}
\end{figure}

\begin{figure}[!h]
\centerline{\includegraphics[width=1.05\columnwidth]{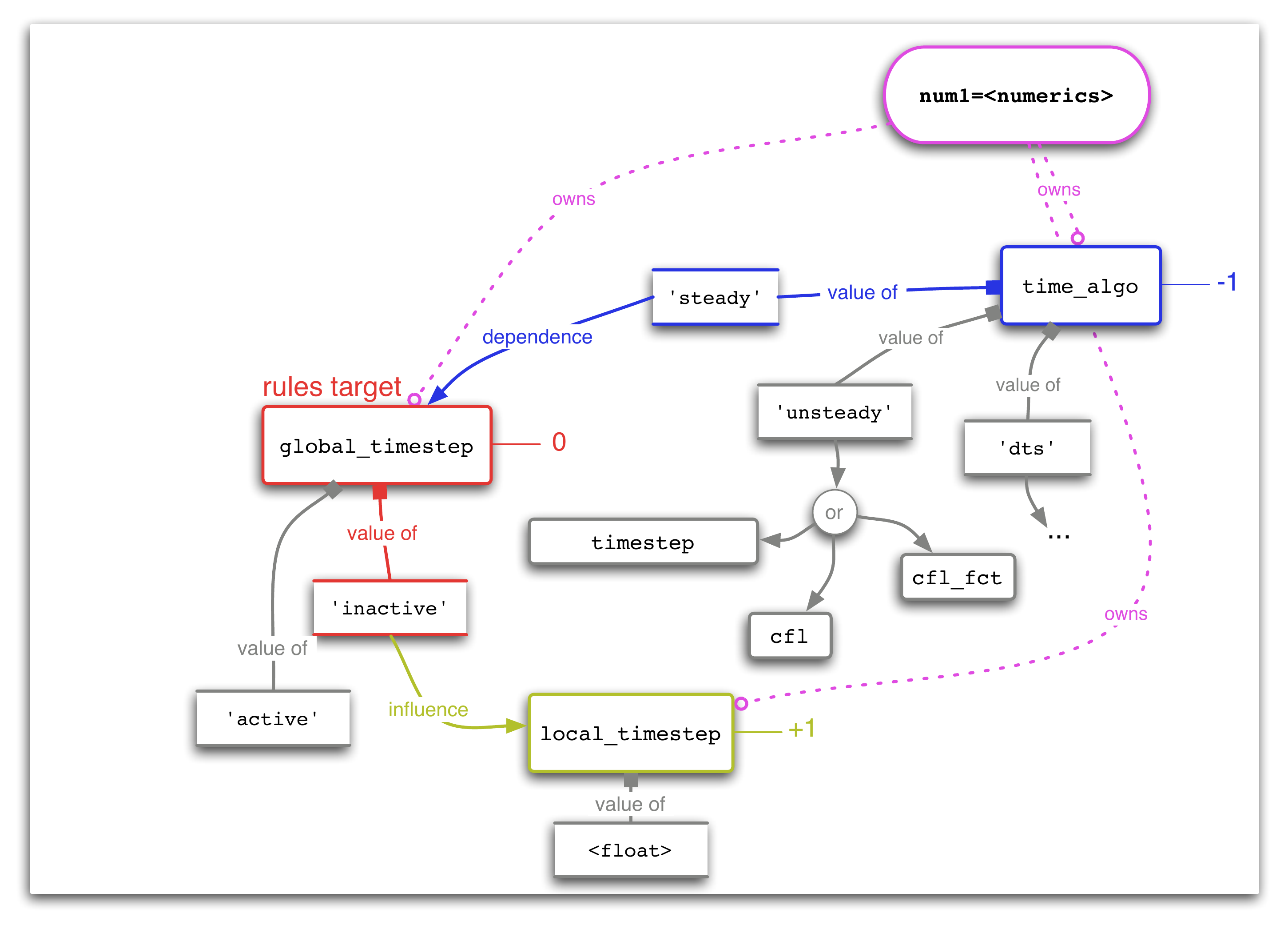}}
\caption{\label{global_tree}Global rules traversal (more than one attribute).}
\end{figure}

Each rule is centered on one attribute, \figurename~\ref{local_tree},
ignoring for the moment the potential complexity which results from
traversing all the rules, \figurename~\ref{global_tree}. When this
traversal is performed for a specific case (with user-specified values
as initial conditions), a dynamic \sigle{DAG} based on contextual
relations (context \sigle{DAG}) is realized. This \sigle{DAG}
typically has many levels, quite differently from the flat and static
one defined by the class structure, which is used for data storage
(and documentation).

Optionally, meaningless values may be ``pruned'' from the context
(using the \comput{<desc>.check(prune=True)} call), hopefully making
it both complete (computable) -- using default values -- and coherent
(here meaning minimal).A simple pruning example would be removing the
choice of turbulence model if the simulation is declared laminar, as
opposed to a plain \comput{<desc>.check()} which would only flag the
combination as non-coherent.

Pruning may not be fully automated, because the rules system cannot
decide on which side of the rule the error lies. Thus, in the spirit
below, we will not make a decision to remove the value: "In the face
of ambiguity, refuse the temptation to guess." \citep{Peters2004}.
\smallskip

For example, the user may have kept the turbulent model definition
from a previous case, or forgotten to change the flow model from
laminar to turbulent when choosing the turbulence model; in this
occurrence, a \comput{check()} call on the involved \comput{model}
object would lead to a \comput{Warning}-level message, see
\ref{messages}.

When performing pruning, the decision is made according to dependency
rules, this is why in the above example the turbulent model choice
would be removed while the flow model (higher up the context
\sigle{DAG}) is left unchanged. On the other hand, if unneeded values
are defined and the \cmdopt{strict} \sigle{CLO} is used (or if
the corresponding internal option is set from the script), the
simulation will be aborted because in this condition all
\comput{Warning}s are bumped to the \comput{Error} level.

In propagating an initial context of user-defined values using the
various contextual rules, we will call ``down-propagation'' context
flow that is centrifugal (propagating away from the root of the
context \sigle{DAG}), and ``up-propagation'' context flow that is
centripetal, see \figurename~\ref{tree_dep_inf}. A \emph{context}
check may start at any point in the \sigle{DAG}, performing both
up-propagation (for dependency) and down-propagation (for influence);
this is useful for the \sigle{GUI} and for interactive user help (what
if).

Example rules will be provided for the \comput{model} class of
\elsAPython/, which gathers (as attributes) parameters related to the
physical model.

\subsubsection{Dependency rules}

A dependency rule defines the up-propagation of an attribute, and
specify that other (source) attributes must have specified values for
coherency, see \figurename~\ref{tree_dep_inf}, e.g.:
\newline
\comput{'visclaw': \{'phymod': ['nslam', 'nstur']\}}
\newline
meaning that the \comput{visclaw} attribute has meaning only for the
\comput{'nslam'} and \comput{'nstur'} (laminar and Navier-Stokes
respectively) fluid models, and not for the remaining \comput{'euler'}
model, see \ref{statdefs}.

\subsubsection{Influence rules}

An influence rule defines the down-propagation of an attribute value,
and specify that other (target) attributes must be defined for
coherency, see \figurename~\ref{tree_dep_inf}, e.g.:
\newline
\comput{'visclaw': \{'sutherland': ['suth\_const', ['suth\_muref', 'suth\_muref\_fct'], 'suth\_tref']\}}
\newline
meaning that the \comput{'sutherland'} value of the \comput{visclaw}
attribute requires that the \comput{suth\_const} and
\comput{suth\_tref} attributes be defined, together with one of
\comput{suth\_muref} and \comput{suth\_muref\_fct}.

So-called ``strong'' influence rules specify moreover that the value of
the target attribute must belong to a specified list, depending on the
value of the origin attribute, e.g.:
\newline
\comput{'user\_config': \{'limited': [\{'turbmod':['keps', 'komega']\}, 'easy']\}}
\newline
would be a strong rule specifying that when
\comput{user\_config='limited'} the only possible choices of
turbulence model are \comput{'keps'} and \comput{'komega'}, and that
the \comput{'easy} attribute value is required.

\subsubsection{Contextual default rules}
\label{contdefs}

Contextual default rules provide a mechanism for defining default
values of attributes when neither user-defined nor static defaults are
provided, e.g.:
\newline
\comput{'suth\_muref': \{1.78938e-5: \{'mixture': ['air'],'cfdpb.units':['si']\}\}}
\newline
meaning that the default value of the \comput{suth\_muref} attribute
is \comput{1.78938e-5} (in \sigle{SI} units), provided that the fluid
composition is defined as \comput{'air'} (from the single-element
\comput{['air']} list) and the problem is specified in \sigle{SI}
units.

The \sigle{DAG} may be dynamically extended through contextual default
rules, which are applied iteratively on a \comput{check()} call until
no new values are defined (or until the maximal iteration count is
reached). For each descriptions context state -- user-defined values,
``static'' default values, and current contextual defaults --
influence and dependency rules, together with ``always required''
rules, define which remaining attribute values must be defined to
render the context complete. Each newly defined value corresponds to a
state transition of this checking automaton.

For all description classes, trade-specific rules (possibly using
regular expressions) may be triggered using the \comput{user\_config}
attribute with arbitrary \comput{<string>} values.

All attribute values may be traced to their creator (kernel, user,
static default, contextual rule) using the \comput{show\_origin}
method. If the creator is a contextual rule, it is listed.

\subsubsection{``Horizon'' mechanism}
\label{horizon}

An original ``horizon'' mechanism has been developed
\citep{Lazareff2009}, replacing the definition of various user skill
levels with a movable by-attribute boundary across the context-defined
\sigle{DAG}; it still has to be documented and user-tested.

\section{Documentation \& error management}
\label{doc}

\subsection{Integrated documentation}
\label{intman}

The \comput{man()} function of the interface provides, for any element
of the \elsAPython/ interface (function, class, method, attribute), a
compact (and always up-to-date) documentation suitable for a returning
user:

\begin{verbatim}
man(check)
Name       : check
Type       : function
Description: Check status of root script object

man(view)
Name       : view
Type       : function
Description: Facade for current script's method

man(model.view)
Name       : view
Type       : instancemethod
Description: Filtering view for a description

man('phymod')
1) Attribute name: phymod
2) Class(es)     : model
3) Description   : fluid model
4) Allowed values: 'euler', 'nslam', nstur'
5) Rules         : 
  5b) influence rules:
    phymod = 'nslam' requires:
      value(s) for visclaw & prandtl & trans_mod & ...
    phymod = 'euler' requires:
    phymod = 'nstur' requires:
      value(s) for visclaw & cv & prandtl &  ...
  5c) context-dependent default values:
    phymod = 'nstur' IF:
      user_config = 'test::wing' | 'test::body' |  ...
  5d) absolute rules:
    attribute value is always required
6) Default value(s): 'euler'
    context-dependent default values in
    '5c)', if any, are applied first
\end{verbatim}

Notice: some output lines have been truncated.

When the same attribute name is shared by several classes, the output
of \comput{man()} is factorized across identical definitions.

\subsection{User's Reference Manual updating}
\label{urm}

A \sigle{PDF} version of the User's Reference Manual (\sigle{URM}) is
built using the {\LaTeX} typographical software \citep{latex}. For
each new version of the \elsA/ software, the \sigle{URM} must be
updated with new attribute descriptions (and possibly new functions,
classes, methods, and additions to the specialized appendices).

The updating process is partly automated, but requires some human
intervention: the \comput{CheckDefs} tool included in the Python
interface code provides automatic generation of the {\LaTeX} source
code for the description of the attributes missing in the current
\sigle{URM} version, based on the contents of the resource
files. Merging in the new {\LaTeX} source is done manually -- using
the \comput{ediff} tool in the \comput{emacs} editor -- and coherency
of the manual with the interface definitions checked, still using
\comput{CheckDefs}.

\subsection{Exceptions (error management)}

The \elsAPython/ interface defines its own exception classes, which
use two severity levels:

\begin{itemize}
\setlength\itemsep{0ex}

\item \comput{WARNING}: non-fatal error;

\item \comput{ERROR}: fatal error.

\end{itemize}

Notice: the \cmdopt{strict} \sigle{CLO} upgrades all
\comput{WARNING}s to \comput{ERROR}s.

\subsection{Error messages}
\label{messages}

The structure of \elsAPython/ error messages is:

\begin{itemize}
\setlength\itemsep{0ex}

\item first line: general information about the error, including the
  severity level;

\item a few lines describing in more detail the problem leading to the
  error;

\item last line: the suggested correction.

\end{itemize}

It has to be noted that, although these messages on their own appear
to be quite explicit, users of our software will sometimes \emph{not
  read} the message, instead ringing up \elsA/ software support.

This does not appear to be uncommon \citep{UX2011}, and could be
linked to user stress, which is what we tried to avoid in the first
place, or to a desire not to do a mental ``context change'' by trying
to actually solve the problem themselves: ``What they want is to pick
up the phone, make a call, and have someone tell them what to do.''
\citep{slash2010}.

\section{The \PyGelsA/ \sigle{GUI}}
\label{pygelsa}

The \PyGelsA/ \sigle{GUI} (Graphical User Interface) is built
on-the-fly using the ``static'' interface definitions, see
\ref{statdefs}; this means that no modification of the \PyGelsA/ code
is needed when updating the user interface for new versions. When
loaded with a script, the \sigle{GUI} then uses the ``dynamic''
definitions, see \ref{dynamic}, for coherency in the display of the
description objects.

\begin{figure}[!h]
\centerline{\includegraphics[width=1.05\columnwidth]{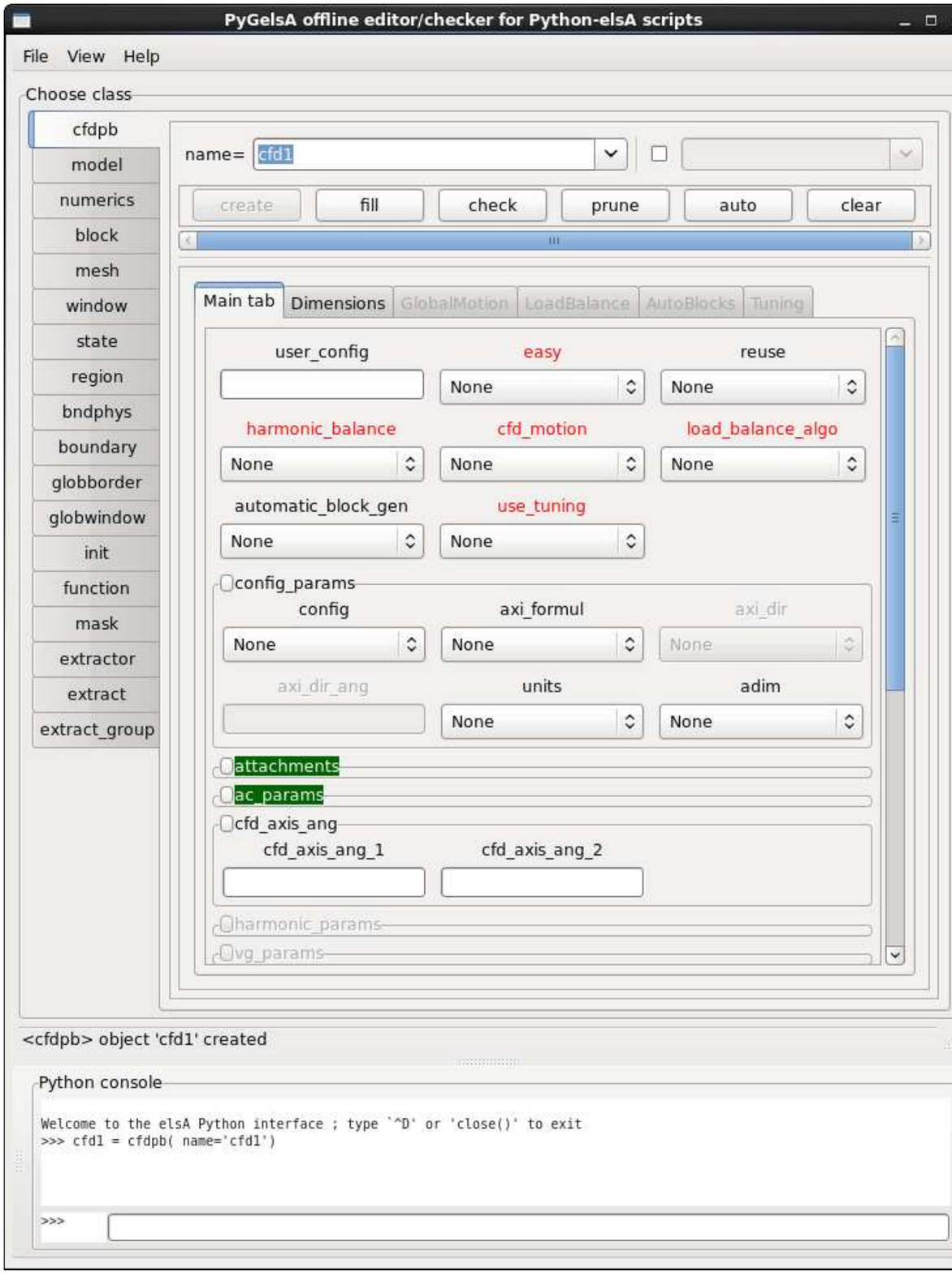}}
\caption{\label{Create_cfdpb_stars}The \PyGelsA/ \sigle{GUI}.}
\end{figure}

This display stays coherent thanks to on-the-fly application of
context checks. As shown on \figurename~\ref{Create_cfdpb_stars},
attributes with missing required values are labeled in red, while
meaningful but user-folded macro-attributes are labeled in green.

The \sigle{GUI} includes (bottom part) a console for text-driven
interaction, which is synchronized with the graphical window
above. The interface state may later be re-built, either from the
logfile or from the file dumped from a \comput{dump()} call performed
at any time during the \sigle{GUI} use.

\section{Database and network operations}
\label{datanet}

Script (and description) objects may be serialized (transformed into
bytecode) and either dumped to/loaded from a database, or accessed
through the network.

\subsection{Database operations}
\label{database}

The script database feature of the interface is built upon a standard
underlying database implementation, which may be as simple as a
``dictionary on disk''.

A script is an in-memory structure of description objects -- and
usually some operations-- built either from an explicit script file or
from code. These operations may be kept ``pending'' (not executed) if
the script is to be stored in a database.

Script databases provide the basic dump/load operations, through the
\comput{dump()} and \comput{load()} methods. A dumped and re-loaded
script will re-create its description objects, and will execute its
pending operations, if the environment (mesh files \ldots) is
adequate.

\begin{figure}[!h]
\centerline{\includegraphics[width=1.05\columnwidth]{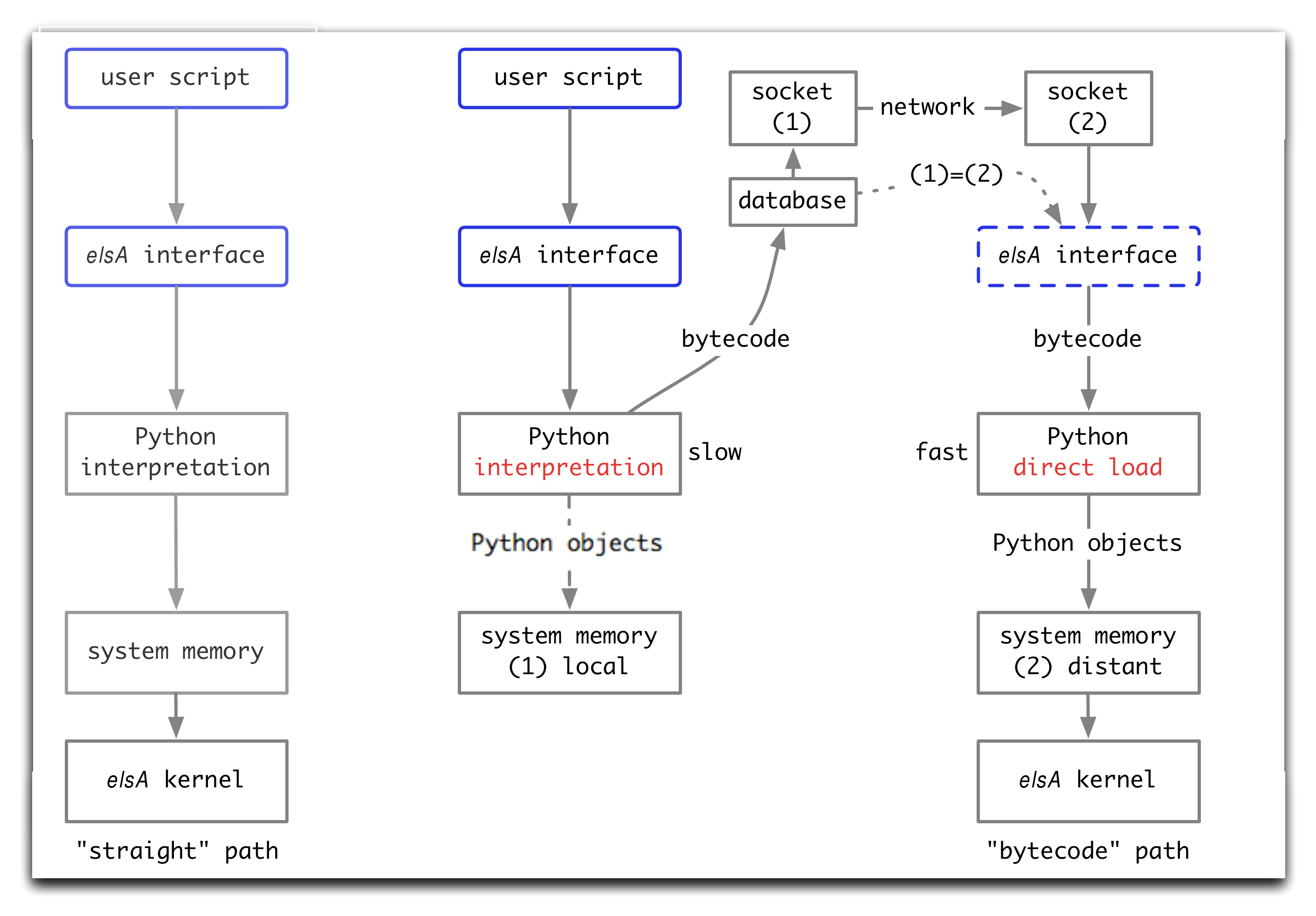}}
\caption{\label{ValDatabase}Database and network use.}
\end{figure}

These basic operations are complemented by a \comput{search()}
method. For the search to be useful and efficient, the user has to
declare a \comput{view}, defining which parameters (out of possibly
several hundred) are of interest for the current study. The
\comput{<scri>.catalog()} method is then called -- using this
\comput{view} -- at each \comput{dump()} to update the database's
catalog. The \comput{catalog()} call may also be used for tagging the
output data for the corresponding simulation, providing an authentic
traceability (e.g. for plots).

Adequate locking allows the database to be shared in read/write mode
between several processes, possibly on different machines, allowing
natural parallelism. The default behaviour is to use separate
databases for script definition (read-only) and for job execution
metadata (read/write) for better efficiency.

The database keeps track of not yet started (\comput{NYS}), running
(\comput{RUN}), and completed (\comput{CMP}) state of computational
jobs, which helps recovering from crashes, generally using a global
\comput{<database>.clean()} call (resetting all \comput{RUN} jobs to
\comput{NYS} state) before restart. This partial restarting has been
very efficient to reduce user stress, especially when running
simulations with more jobs than computing nodes and (clock) execution
times counted in days.

\subsection{Network operations}
\label{network}

Methods for the \comput{network} class include a simple server/client
(sockets) pair, the \comput{dump()}/\comput{load()} pair, and the
\comput{popen()} (get command output) and \comput{db\_search()} (find
in database) commands.

This functionality allows to access a script database through the
network, which is useful when no shared filesystem is available and
the underlying database is not networked, see
\figurename~\ref{ValDatabase}.

\section{Extensibility}
\label{extend}

\subsection{Products}
\label{products}

So-called ``products'' are additional (Python) software packages which
couple with \elsAPython/ (and possibly with other products) through
description classes. A developer may build a new \comput{product}
(based on the provided skeleton) and simply drop it in the
\comput{Products} directory. Communication between \elsA/ and the
added software is performed through an instance of this
\comput{product}-based description class, which contributes to the
global context and may get to steer the simulation in lieu of the
\sigle{CFD} solver, see section \ref{doe}.

The algorithm associated with the product may be coded in Python
(possibly referencing external packages with compiled libraries)
and/or use \elsA/ kernel functionality.

\subsection{\computn{root} and \computn{boot} objects}
\label{rootboot}

If a description class declares a \comput{compute()} method, an
instance of this class may grab control of the simulation from the
\sigle{CFD} algorithm, using the global context to gather all the
required data. This instance is then called the \emph{boot object}.

Scripts define a \comput{compute()} method -- invoked in the
(top-level \comput{root}) script through the \comput{compute()}
function -- which is an indirection to that of the current
\comput{boot} object; a script with a \comput{compute()} statement
will thus execute differently -- without explicit tests -- depending
on the current boot object, or on the explicit
\comput{<desc>.compute()} call, e.g. (see \ref{sfddmdspi}):

\begin{verbatim}
cfd1 = cfdpb(name='cfd1')
cfd1.set('sfd', 'active')
...
dmd1 = dmd(name='dmd1')
spr1 = sparse_poly(name='spr1')
...
# cfd1.compute() # single-point SFD simulation
# dmd1.compute() # single-point SFD/DMD simulation
compute() # SFD/DMD/SPI coupling on DoE, using spr1
\end{verbatim}

\noindent where \comput{\#} starts a comment, and
\comput{compute()} is redirected to \comput{spr1.compute()} because
\comput{spr1} is the last-created bootable object.

This logic provides for testing a complex coupling script at different
levels with minimal modification; it can also be implemented as:

\begin{verbatim}
...
slvrs = {0:cfd1, 1:dmd1, 2:spr1}
slvr_lev = 1 # use dmd1
slvrs[slvr_lev].compute()
\end{verbatim}

The coupling level may thus be specified as an integer value, without
any test in the user script.

Lastly, the \comput{slvrs[slvr\_lev].compute()} call above may be
replaced by \comput{set\_boot\_objt(<desc>); compute()}, where the
first statement may be seen as passing a \emph{token} to
\comput{<desc>}, giving it specific rights. This token may be moved
during the simulation, i.e. for collaborating coupling algorithms.

\subsection{\computn{provide()} method for blind creation}
\label{provide}

When several \comput{product}-like description objects
(e.g. \comput{prd1}, \comput{prd2}) use a common description object
(e.g. \comput{slave}), it may occur that it is built by one of them,
which should be the first appearing in the script. If for example
\comput{slave} is built by \comput{prd1.compute()}, built first, and
then used by \comput{prd2}, all is well. But if the \comput{boot}
object is switched from \comput{prd1} to \comput{prd2}, \comput{slave}
will not be built and will miss to both \comput{prd1} and
\comput{prd2}, as \comput{prd1.compute()} is not called anymore.

The solution we use, to avoid here an explicit test on the existence
(with possible creation) of \comput{slave} before each reference, is
the use of the \comput{provide()} method, which wraps the conditional
creation, and may be ``blindly'' called to always return the (new or
pre-existent) object.

\subsection{\computn{target\_lift} class for target lift computations}
\label{targetlift}

The ``target lift'' functionality allows to replace a standard
lifting-body computation, at fixed angle-of-attack (\AoA/), with the
computation of the list of \AoA/s corresponding to the given list of
lift values (within convergence bounds).

This is performed by replacing the standard \comput{compute()} call
with one directed to a new \comput{<target\_lift>} boot object, as in:
\begin{verbatim}
tcl1 = target_lift(name='tcl1')
tcl1.attach(lift)
alphas = compute([0.05, .10, .15])
\end{verbatim}

where \comput{lift} is an \comput{<extractor>} defining the
computation of the body's lift and \comput{alphas} is the \AoA/
output for the specified lift coefficients.

This was the first example of an extension to be finally coded as a
\comput{product}.

\subsection{\computn{variator} class for parametric studies}
\label{variator}

The \comput{variator} class allows building script variations from a
base version and a list of parameter perturbations, dumping them to a
database. Later on this database will be spanned, meaning that all the
database scripts will be loaded and pending operations, see
\ref{database}, executed. This is performed in sub-directories of a
user-chosen base directory, with all file paths automatically shifted
as required. Spanning thus provides the database with an
\emph{iterator}.

A step further, this iterator is made into a more general
\emph{automaton}: during spanning, a simulation may be restarted
(chained) from a source simulation, chosen using a user-defined
non-isotropic \DoE/ distance (in parameter space), possibly with
additional rules for restricting the source/target pairing. When
chaining simulations, the automaton may also add intermediate points
(linearization) to the iteration list, to respect a user-specified
maximal jump size in parametric space \citep{Lazareff2014}.

An external algorithm may also add arbitrary points to the list, see
\ref{sfddmdspi}.

\subsection{\computn{swarm} class for operational efficiency}
\label{swarm}

The \comput{swarm} class provides an abstraction for a group of
spanning simulations. For operational efficiency, automated load
management on a computational node is provided either as a maximal
count of swarm job instances or as a fraction of the node's power.

The values of the selected observable quantity, for all the
simulations in a swarm, are returned as \comput{<valu\_list> =
  <swarm>.compute()}, as used for the \sigle{SPI} algorithm,
\ref{sfddmdspi}.

\section{Design of Experiment studies}
\label{doe}

Studies in Design of Experiment (\DoE/) space involve at least two
levels: the basic automaton performing the space spanning, and the
mathematical tools, applied both to each \DoE/ point and to the global
space. These tools may be passive (using specified \DoE/ points in a
given order), or they may use observable quantities to steer the
spanning automaton, possibly creating new \DoE/ points on-the-fly.

In \ref{sfddmd} and \ref{sfddmdspi} we will give two examples in the
field of differentiable dynamical systems (the geometrical study of
complex systems and their stability) \citep{Smale1967} applied to
\sigle{CFD}, first as a local study and then on a \DoE/. Usually the
system is defined by an equation like $\dot{x} = v(x)$. Here however
the operator of the system is unknown, and only accessible through
\sigle{CFD} observable quantities at computed points of the \DoE/, so
that as a first step we have to address the problem of \DoE/
discovery, see \ref{span}, to ensure that the main features of the
operator are accounted for.

\subsection{\DoE/ spanning and discovery}
\label{span}

\DoE/ discovery is an extended parametric study, see \ref{variator},
where the aim is to describe (find the main features of) the space
inside a closed boundary in parametric space, with possible refinement
of the initial set of \DoE/ points \citep{Lazareff2014}. Continuation
techniques (e.g. stabilization to cross \DoE/ areas with separated
flow) may be needed, see \ref{sfddmd}. As a first approach, \DoE/
refinement may be manual, but in \ref{sfddmdspi} we introduce an
automatic refinement algorithm through the \sigle{SPI} algorithm
\citep{Chkifa2014}, which greatly simplifies user interaction.

The observable quantity here may be the aerodynamic lift or drag, the
spectral radius of the global (physical + numerical) operator, or any
other quantity of interest.

\subsection{Coupling with the \sigle{SFD} and \sigle{DMD} algorithms}
\label{sfddmd}

Coupling between \elsA/, the \sigle{SFD} and \sigle{DMD} algorithms
has been performed to provide stabilized \sigle{CFD} solutions,
initially at a single \DoE/ point for the flow around a cylinder
\citep{Cunha2015}. The \sigle{SFD} algorithm may be either part of the
\sigle{CFD} solver or treated as a wrapper in the encapsulated version
\citep{Jordi2014}, with an \comput{sfd} description class. The
\sigle{DMD} algorithm is coded in Python, with a \comput{dmd}
description class. Some parameters of the \sigle{CFD} algorithm
(\comput{numerics} class attribute values) are dependent on the
\sigle{SFD} and \sigle{DMD} algorithms through contextual rules.

This work has since been extended to a (Mach, Reynolds) \DoE/ using
the \comput{variator} class. Success is still dependent on the
adequate adjustment of some parameters, at this time taken to be
constant across the \DoE/. For the time being, this aspect is fully
dependent on user expertise.

\subsection{Coupling with the \sigle{SFD}, \sigle{DMD}, and \sigle{SPI} algorithms}
\label{sfddmdspi}

The most complex application to date of the extensibility of
\elsA/ through the \elsAPython/ interface is the algorithm
described above in \ref{sfddmd}, complemented by Sparse Polynomial
Interpolation (\sigle{SPI}) \citep{Chkifa2014} for \DoE/ discovery (to
be published).

Here the spanning automaton is managed by the \sigle{SPI} algorithm,
starting with nodes only at the \DoE/ summits and progressively
enriching the representation with inner points according to a
Clenshaw-Curtis distribution.

The \sigle{CFD} results for the chosen observable quantity at each
refinement level of the \sigle{SPI} algorithm are computed as a swarm,
see \ref{swarm}.

With a reasonable degree of convergence of the \sigle{SPI} algorithm,
we hope that this algorithm will lead to an adequate automatic
discovery of the \DoE/ features.

\section*{Conclusion}

The basic foundation of the \elsAPython/ interface, which has been
used since the creation of the \elsA/ software, has long been used for
the target lift extension, see \ref{targetlift}. Recently more complex
extensions have been introduced for \DoE/ studies.

This descriptive interface and associated functionality provides a
high level of abstraction for coupling mathematical algorithmic layers
with the \sigle{CFD} solver, reducing the tedious part of user's work
and augmenting efficiency in research and applications.

This approach, allowing for collaborative computing, seems to be well
suited for projects coupling several simulation kernels.

\section*{Perspectives}


A rule-based implementation of the currently kernel-side factory, see
\ref{base} would remove the potential fragility of the current one,
where tests are sometimes nested for several levels. A rule only has
to be written once, while inner tests may have to be repeated for each
branch of the upper level(s). This would benefit to user-side checking
by ensuring completeness and coherency of the rules with the actual
solver implementation.

The more immediate new development will be stochastic analysis, using
the Monte-Carlo method on the response surface of \ref{sfddmdspi},
introducing the first 4th-level layer algorithm above the \elsA/
interface (5th-level above the \sigle{CFD} kernel).

\tableofcontents

\section*{Acknowledgments}

This work has received partial financial support from the Agence
Nationale de la Recherche (\sigle{ANR}
\url{http://www.agence-nationalerecherche.fr/}), first through the
Carnot institutes network contract referenced as n\textsuperscript{o}
07 CARNOT 00801, and then as part of the \sigle{UFO} (Uncertainty For
Optimization) project referenced as ANR-11-MONU-0008.

\medskip
This paper is dedicated to my late colleague and friend Michel Gazaix,
who died unexpectedly on August 8, 2016. He was the one, as software
architect, who invited me into the early \elsA/ project to design and
build its user interface, and I am still indebted to him for this
challenge.

\section*{Bibliography}

\RaggedRight 
\bibliographystyle{elsarticle-harv}
\bibliography{biblio.bib}

\end{document}